# Unravelling metallic contaminants in complex polyimide heterostructures using deep ultraviolet spectroscopic ellipsometry

Muhammad Avicenna Naradipa[1,2], Prayudi Lianto[3], Gilbert See[3], Arvind Sundarrajan[3], Andrivo Rusydi[1,2,4,*]

[1]Advanced Research Institute for Correlated-Electron Systems (ARiCES), Department of Physics, National University of Singapore, 2 Science Drive 3, Singapore 117551, Singapore

[2]Singapore Synchrotron Light Source, 5 Research Link, Singapore 117603, Singapore

[3]Applied Materials, 10 Science Park Road, Singapore 117684

[4]NUS Graduate School for Integrative Sciences and Engineering, National University of Singapore, Singapore 117456, Singapore

*Corresponding Author: andrivo.rusydi@nus.edu.sg

# Abstract

Metallic contaminants in complex heterostructures are important topics due to their significant roles in determining physical properties as well as devices' performance. Particularly, heterostructures of polyimide via on Al pad and Cu redistribution layer (RDL) on polyimide have shown exotic properties and are important for advanced semiconductor packaging systems. One main problem is significant leakage current variations, which affect the performance of the devices, yet its origin is far from understood. Furthermore, metal contaminations, *if any*, would occur at the buried interfaces and it is particularly challenging to probe them. Until now, the electronic and optical properties of complex polyimide heterostructures and the roles of metallic contaminants have not been studied extensively. Herewith, using spectroscopic ellipsometry (SE) in broad deep ultraviolet (DUV) range supported with finite-difference time-domain (FDTD) calculations, we determine optical properties with various concentration of contaminants and their influence on device performance of under-bump vias and redistribution layer (RDL) architectures, especially at the metal-bump interface and surface between RDL. The complex dielectric function reveals varying contamination levels and different metals responsible for chip performance. Metallic contaminants are embedded within ~50 nm in the polyimide and different metals are distinguishable with varying concentrations (1.3% – 30% relative volume fraction), in agreement with contact measurements in highly complex structures. Our result shows the potency of spectroscopic ellipsometry in the DUV and paves the way for non-destructive, advanced quality control and metrology applications in integrated advanced electronics packaging systems.

With today's increasing use of microchips on mobile devices, the Internet of Things, and high-performance computing systems, chip manufacturers are demanded to innovate and fabricate chips with low cost, compact size, and a high number of connections (pin count). Advanced electronics packaging systems play an important role, as it determines the quality and efficiency of microchips to the metal circuitry and printed circuit board (PCB). One of the most common methods to connect microchips to the PCB is through so-called fan-out wafer-level packaging (FOWLP), where a metallic layer (redistribution layer, RDL) is used to distribute ('fan out') connections between different components of the PCB[1,2]. In FOWLP configuration, RDL removes the necessity of organic substrates, which has been shown to improve signal integrity, thermal dissipation, maximum junction temperature, and component density[3]. This allows the fabrication of application-specific packaging and system integration, highly preferred for mobile applications[2,3].

Quality control during RDL fabrication requires *in-situ* non-destructive techniques to directly probe surface and interface contaminants. However, it was challenging to probe such contaminants as several issues arise during fabrication, such as cracks (undercut), bends, (warpage), and current leakage caused by metal residue contamination[4-6]. Fine-tuning fabrication conditions and process flow to counter these issues are non-trivial, as they are intertwined and correlated with one another[2,7]. The metal contaminations, for instance, pose a significant issue as the induced leakage current varies up to seven orders of magnitude[2,5]. Moreover, contamination on the metal bump pad (connection between the chip and PCB) have shown increased contact resistance ($R_C$), reducing the performance and durability of the chip[7]. Since the metal residue is embedded in the RDL or bump pad, it is challenging to probe such a contamination. The contamination detection is traditionally conducted using *ex-situ* x-ray photoemission spectroscopy[2] (XPS), fluorescence and transmission microscopy[8], and scattering metrology[9]. Other *ex-situ* microscopy methods are also highly utilized, such as atomic force microscopy (AFM) and scanning transmission electron microscopy (STEM). Unfortunately, these characterization techniques are generally destructive and cannot observe underlying electronic and optical structures due to contaminants. Moreover, X-ray-based characterization requires high brightness, thus difficult to integrate into compact, *in-situ* metrology systems due to long measurement times[10].

Ellipsometry has been implemented to study semiconductor wafers and electronics packaging, such as thickness determination[11], and optical properties of 3D semiconductor devices, nanosheet fin structures, and 3D interconnects[9,12-14]. Recent studies have also shown its surface-sensitive capabilities on polymers in the IR range[12,15]. Due to interference effects, metallic contaminants, however, are difficult to observe in the IR range. The absorption of other components near visible-UV range also complicates analysis in devices with stacked layers and complex microstructures. Therefore, a different approach is needed to reveal electronic and optical properties of complex polyimide heterostructures.

In this Letter, we reveal the electronic and optical properties of polyimide via on Al pad and Cu RDL on polyimide with varying metallic contamination levels in advanced semiconductor packaging systems using a non-destructive characterization method of spectroscopy ellipsometry (SE) in a broad energy range, up to deep ultraviolet (DUV). Supported with theoretical finite-difference time-domain (FDTD) calculations, the complex dielectric function shows significant changes in DUV due to electronic correlations, allowing SE to identify the type and concentration of buried metallic contaminants.

First, we first fabricate blanket polymer (polyimide) on Si substrate with varying metallic contamination levels (Fig. 1a). The polyimide contains a controlled level of Ti and Cu contaminants, ranging between 1 – 9 % and 0 – 20% for Ti and Cu, respectively. The XPS uses a monochromatic Al Kα (40 W, 15 KeV) source with a spot size of 100×100 μm. The pass energy is 55 eV with an energy resolution of 0.1 eV. The depth of metallic contaminants is determined using *ex-situ* XPS continuous sputter/etching mode with Ar ions. The percentage represents the relative volume fraction of the contaminants embedded in the polymer.

Next, we fabricate two patterned test vehicles, to address two key challenges in terms of interface/surface cleanliness. The first test vehicle measures contact resistance ($R_c$) to characterize cleanliness of Al-Ti interface. $R_C$ test vehicle is fabricated using the process flow described in Ref.[7], with some modifications (Fig. 1b). Standard microbump process flow is conducted to form passivation layer encapsulation around Al pad peripheries. To generate varying levels of contamination, pre-clean splits are introduced at the under-bump metallization (UBM) deposition step. Contact measurements are conducted using four-point-probe Kelvin structure to characterize $R_C$, as described in Ref. [7]. The second test vehicle (Cu Redistribution layer, RDL) is fabricated using the process flow described in Ref. [2]. Standard RDL process flow is conducted to form an isolated Cu RDL array on the polyimide surface.

Page 5 of 20

Spectroscopic ellipsometry (SE) measurements are conducted using a custom-designed system to provide ultra-high vacuum compatible, broadband, high-resolution, variable-angle spectroscopic ellipsometry in the photon energy range up to DUV. Note that our *integrated* SE covers photon energy range from 0.01 to 6.5 eV and, together with high energy reflectivity, covers a high-energy range up to 1500 eV[16,17]. The SE measurements are performed at room temperature with three angles of incidence in reflectance mode, $\theta_i = \theta_r = \theta$ and $\theta = 50°, 60°, 70°$ (Fig. 1a), with a beam spot diameter of 0.3 cm × 0.8 cm at 70°. For Cu RDL, focusing optics are used to reduce the beam spot diameter to 0.2 × 0.5 mm at 70°. Due to the large beam spot, multiple metal pads and via structures are measured. The detected ellipsometric data is averaged using dynamic averaging. Spectroscopic ellipsometry measures the amplitude ratio ($\Psi(\omega)$) and phase difference ($\Delta(\omega)$) that directly corresponds to the real and imaginary part of the complex dielectric function ($\tilde{\varepsilon} = \varepsilon_1 + i\varepsilon_2$). In the IR – visible region (below 3.5 eV) polyimide is transparent, resulting in visible fringes in the $\Psi(\omega)$ and $\Delta(\omega)$ (see supplementary materials, Figs. S2-S3). We utilize these fringes as a method to estimate the thickness of the polyimide film. The total thickness estimated by SE is ~6 μm, which is in good agreement with fabrication standards. The surface roughness is less than 2 nm as estimated from atomic force microscopy (AFM). Above 3.5 eV, metallic contaminants significantly impact the optical properties; hence, we focus the study on the optical properties at 3.5 eV – 5 eV. Metallic contaminants are modeled as a separate, continuous film layer on top of the polyimide. To extract $\tilde{\varepsilon}(\omega)$, we use a combination of Gaussian, Tauc-Lorentz[18], Herzinger-Johs PSEMI-Tri[19] generalized oscillators and obtain the best fit[19-21]. Optical properties, such as reflectance at normal incidence $R(\omega)$, the real component of optical conductivity $\sigma_1(\omega)$, and spectral weight ($W$) are derived from $\tilde{\varepsilon}(\omega)$. All fit results show a root mean square (RMS) error of less than < 10. Further details are provided in the supplementary materials.

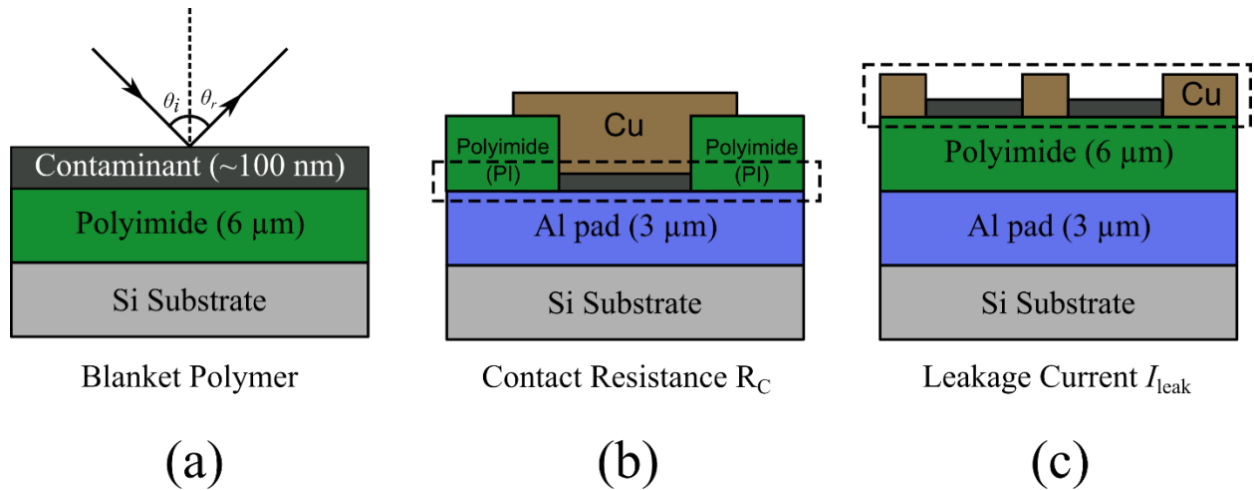

**FIG. 1.** Schematic and Scanning Electron Microscopy (SEM) images of studied test vehicles. (a) Experimental configuration with blanket polymer. (b) Contact resistance ($R_C$) via schematic and (c) Leakage Current test vehicle (Cu RDL) schematic as described in by Lianto et al.[2,7] Spectroscopic ellipsometry measurements of contact resistance ($R_C$) via and Cu (RDL) are conducted with identical experimental configuration in (a).

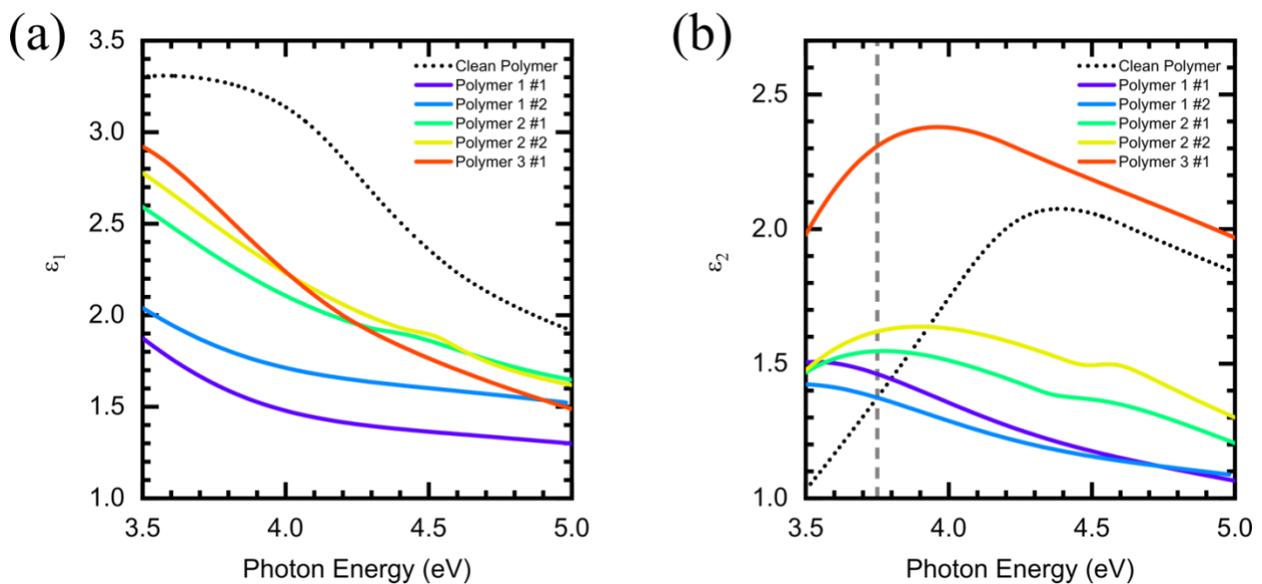

**FIG. 2.** Complex dielectric function $\tilde{\varepsilon} = \varepsilon_1 + i\varepsilon_2$ of blanket thin film polymer on Si. (a) Real ($\varepsilon_1$) part and (b) imaginary ($\varepsilon_2$) part of the complex dielectric function. #1 and #2 indicate production batches with similar fabrication and cleaning processes.

**TABLE I.** Surface Contamination of blanket polyimide on Si determined by XPS.

| Test Vehicle | Ti 2p % at surface | Cu 2p % at surface |
|---|---|---|
| Clean Polymer | 1.3 | 0 |
| Polymer 1 | 7.8 | 10.9 |
| Polymer 2 | 9.8 | 16.3 |
| Polymer 3 | 8.6 | 21.5 |

We conduct SE on the blanket thin film polymer on Si as a control and feasibility assessment. The blanket thin film polyimide is modelled as a polyimide layer based on the optical properties of Palik[22] (Fig. 1a). The contamination layer is modeled as a separate, continuous film containing mixed optical properties of the polyimide and the contaminants (Cu and Ti), located above the 6 μm homogenous polyimide (Supplementary Note 2). Clean polymer is provided as a comparison, where the contamination only reaches 1.3%. The depth of metallic contaminants is ~100 nm from the surface, according to *ex-situ* sputter-mode XPS. We use this as an estimate of the effective contamination layer (Supplementary Notes 1 and 2), setting the thickness at 100 nm for all test vehicles to avoid correlation of parameters. Despite the simple approximation, the optical properties of Ti[23] and Cu[24], which are optically absorbing between 3.5 – 6.5 eV, will influence the electronic and optical properties of the polymer.

In Fig. 2, the optical properties of clean polymer and polymers 1 to 3 are shown. Cleaner samples have lower $\varepsilon_1(\omega)$ and $\varepsilon_2(\omega)$. In particular, Fig. 2b reveals a clear separation of $\varepsilon_2(\omega)$ between highly contaminated (polymer 3) and least contaminated test vehicles (polymer 1). For instance, at ~4 eV, the average $\varepsilon_2(\omega)$ from batch #1 and batch #2 of Polymer 2 are 33.8% lower in comparison to Polymer 3. The $\varepsilon_2(\omega)$ of Polymer 1, the least contaminated sample, is 44.5% lower than Polymer 3. The optical properties of contamination layer, although indicating higher $\varepsilon_2(\omega)$ at 3.74 eV in comparison with the clean polymer, exhibit clear distinction at different contamination levels. We attribute the different optical properties between the Clean Polymer and contaminated polymers (Polymer 1, Polymer 2, and polymer 3) to the distribution of contaminants at the surface and interaction between contaminants and the polymer. To evaluate the validity of the optical model, we include surface roughness based and thickness variation (Supplementary Materials, Fig. S11 and Table S4). When surface roughness is included, the optical properties of the polymers show similar separation of $\varepsilon_2(\omega)$ determined by the concentration of contaminants, where $\varepsilon_2(\omega)$ of Polymer 2 is ~32% lower than Polymer 3. Similarly, $\varepsilon_2(\omega)$ of Polymer 1 is ~49% lower than Polymer 3, indicating the reproducibility and consistency of our optical model (Figs. S12-S13, Supplementary Materials).

To further quantify the SE results, we conduct correlation studies on the complex dielectric function and spectral weight ($W$) with respect to the surface contaminant level detected from XPS (Fig. S4). Based on the extracted $\tilde{\varepsilon}(\omega)$, there is a strong, positive linear correlation for Cu 2$p$ % and $\varepsilon_1(\omega)$, reaching $R^2$ value of 0.86. Meanwhile, the Ti 2$p$ % correlation is lower, which is likely due to its lower relative fraction volume in comparison with Cu. On the other hand, $\varepsilon_2(\omega)$ for both Cu 2p % and Ti 2p % show a positive correlation as their optical properties are similar between 3.5 – 5.0 eV. The spectral weight correlation indicates a similar behavior to $\varepsilon_2(\omega)$, since $W \propto \varepsilon_2(\omega)$. The comparisons show evidence of a strong correlation between the optical properties and the contamination levels. It is expected that SE can detect even lower levels of contamination by comparing with near zero contamination. Based on the correlation study and optical properties, it is clear that SE can detect metal in a broad range and is highly adaptable to different test vehicles, as further discussed below.

We conduct finite-difference time-domain (FDTD) calculations using ANSYS HFSS to qualitatively investigate the reflectance of the contamination layer and the validity of the continuous thin film model used in SE (Fig. 3, see Supplementary Note 4). A 60 × 30 × 30 nm block of polyimide is embedded with a block of Cu and Ti contamination. We vary the depth of the contamination and cross-section area to represent changes in the relative volume fraction (Fig. 3a). The reflectance at normal incidence ($R(\omega)$) of the blanket polyimide is directly attainable from the complex dielectric function using Fresnel equations (Fig. 3b and Supplementary Note 3). Our theoretical calculations (Figs. 3c-f) mimic the reflectance of the contamination layer in the polyimide test vehicles, in agreement with optical properties from SE (Fig. 3b). As the contaminants decrease (shown in percentages in Figs. 3c-f), the reflectance intensity is reduced. In Figs. 3c and 3d, Cu contaminants are varied by changing the horizontal distribution (contamination width) and vertical distribution (contamination height). Although the SE optical model assumes a random distribution of contaminants, FDTD calculations reveal the impact of contaminant distributions. Increasing the contamination height (H = 60 nm, Fig. 3d) reduces reflectivity, but the spectra are more monotonous than H = 50 nm (Fig. 3c). In contrast, Ti contamination increases reflectivity when the contamination level is reduced (Figs. 3e, 3f). Interestingly, by comparing the calculation with H = 60 nm, the calculations qualitatively agree with SE, where the reflectivity profile predicts that the contaminants in polymer 1 to polymer 3 (Fig. 2) are buried approximately 10% – 20% (20 – 50 nm) into the polymer. Thus, FDTD calculations suggest that the concentration of contaminants spreads horizontally for increasing concentrations of Cu and Ti.

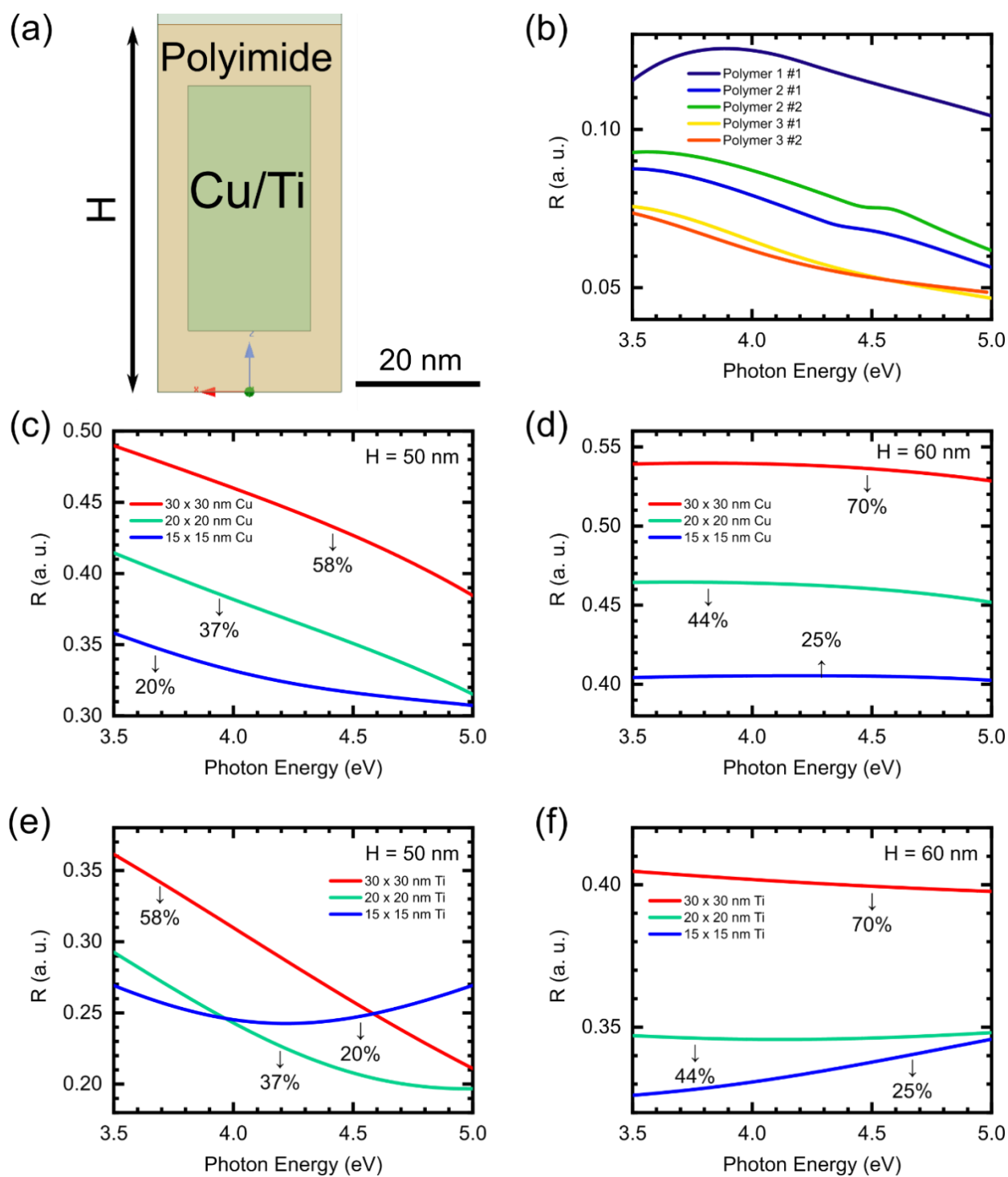

**FIG. 3.** Reflectance at normal incidence $R(\omega)$ from spectroscopic ellipsometry and Electromagnetic Finite-difference Time-Domain (FDTD) calculations from ANSYS HFSS. (a)

Schematic of the FDTD calculation with embedded metal contamination with 50 nm height (labelled as H) and 20% contamination. The polyimide height is 60 nm. (b) $R(\omega)$ calculated from complex dielectric function obtained in SE. $R(\omega)$ calculated using FDTD for embedded (c-d) Cu and (e-f) Ti contamination with contamination height set at 50 nm and 60 nm. The percentage represents relative volume fraction between the contaminants and the polyimide (Cu/Ti:Polyimide).

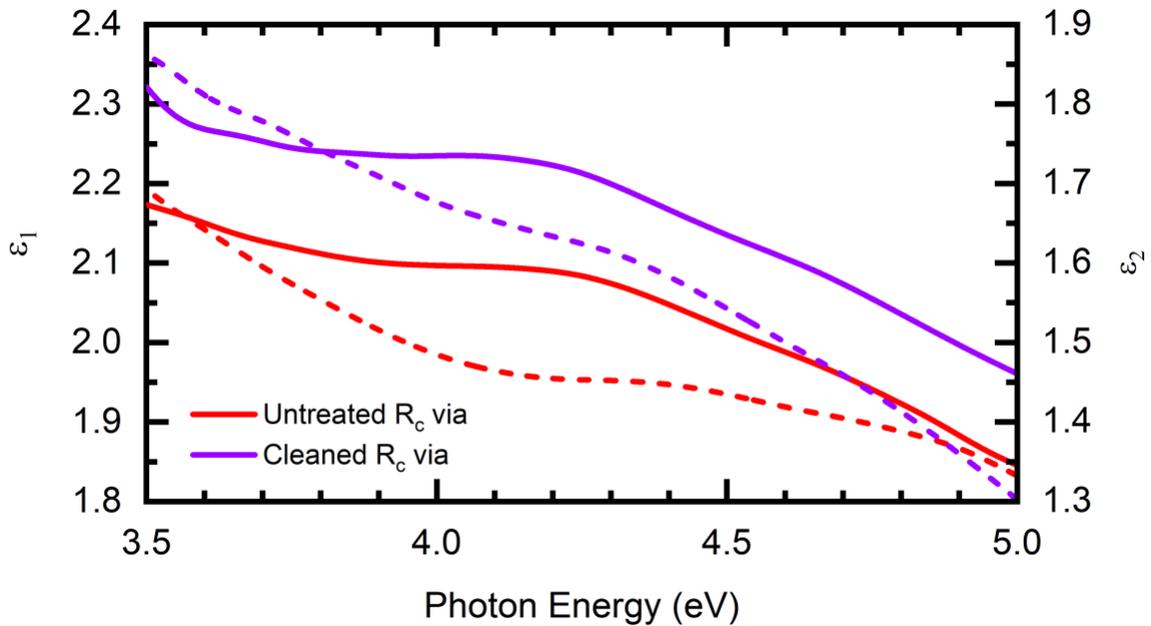

**FIG. 4.** Complex dielectric function $\tilde{\varepsilon} = \varepsilon_1 + i\varepsilon_2$ of polyimide on $R_C$ via. Solid and dashed lines indicate the real ($\varepsilon_1(\omega)$, left axis) and imaginary ($\varepsilon_2(\omega)$, right axis) parts of the dielectric function, respectively.

We next focus on a smaller area of contamination by examining polyimide on $R_C$ via (Figs. 1b and 1d). Via structures are metal connectors between vertical levels of the chip packaging, which are prone to metallic contamination during fabrication. A control (untreated) and a cleaned polyimide test vehicle is processed using AMAT pre-clean chambers. The untreated and cleaned $R_C$ via is measured using Kelvin (4-point) resistance measurement and showed contact resistance of ~$10^2$ mΩ and ~2 mΩ, respectively. Since $\varepsilon_2(\omega)$ is proportional to optical absorption $\alpha(\omega)$ and optical conductivity $\sigma_1(\omega)$, we can correlate the obtained optical properties from SE with complementary contact measurements to inspect each sample's performance. We note that higher levels of contaminants induce higher leakage current through the polymers[2,7], a key parameter in chip performance. In Fig. 4, cleaned $R_C$ via exhibits higher $\varepsilon_2(\omega)$ compared to untreated polyimide throughout the whole energy range. At ~4.0 eV, the difference between untreated and cleaned $R_C$ via is 6.1% and 11.4% for $\varepsilon_1(\omega)$ and $\varepsilon_2(\omega)$, respectively. Despite the simple approximation, SE reveals significant difference between untreated and cleaned samples, signifying the effectiveness of the model at broad wavelengths.

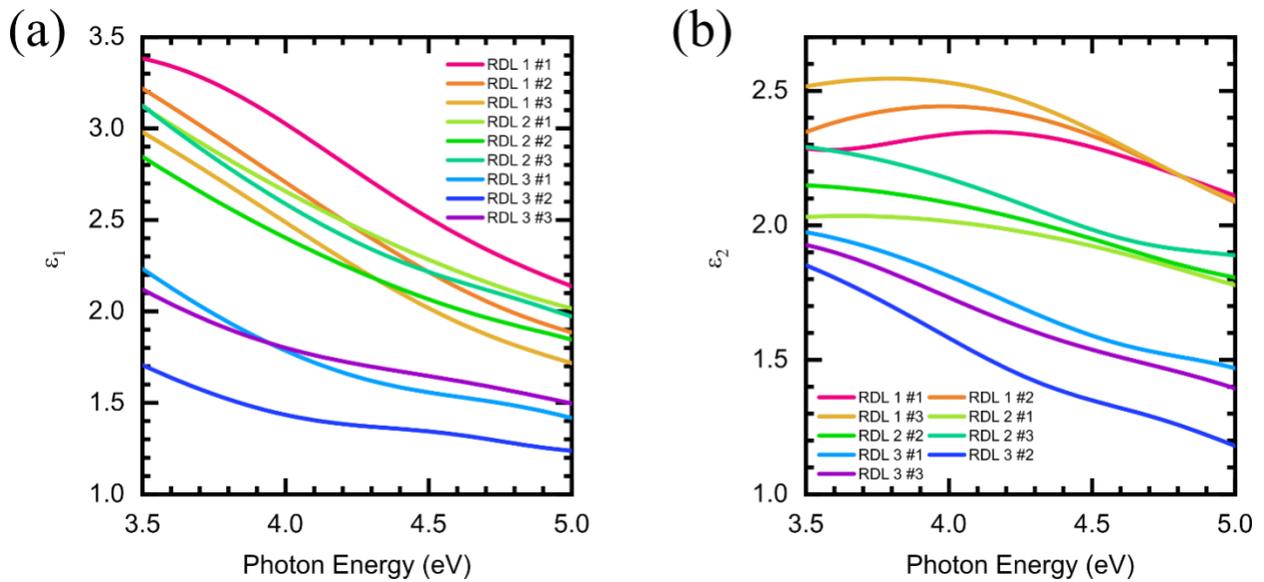

**FIG. 5.** Complex dielectric function $\tilde{\varepsilon} = \varepsilon_1 + i\varepsilon_2$ of Cu RDL. Test vehicles treated with 0-, 10-, and 20-minute cleaning processes, labeled as RDL 1, RDL 2, and RDL 3, respectively. (a) Real ($\varepsilon_1$) and (b) imaginary ($\varepsilon_2$) parts of the dielectric function, respectively. #1 and #2 indicate production batches with similar fabrication and cleaning processes.

We perform SE to identify the current leakage ($I_{leak}$) in Cu RDL, where the surface contains the most complex microstructure (Fig. 1c). The die housing of Cu RDL interconnect is 15 mm x 15 mm, where each die contains multiple interconnects with varying diameters. Two sets of Cu RDL are fabricated with varying contamination levels. The untreated sample (RDL 1), has a high leakage current of ~0.1 A at 10 V. In contrast, the cleaned samples RDL 2 and RDL 3 have a measured leakage current of ~$10^{-9}$ A and ~$10^{-14}$ A, respectively. In Fig. 5, we show the complex dielectric function of RDL 1, RDL 2, and RDL 3. It can be seen that $\varepsilon_2(\omega)$ is proportional to the number of metallic contaminants in the sample, where low $\varepsilon_2(\omega)$ indicate cleaner RDL and reduced $I_{leak}$. Reproducibility tests using another batch (marked #2 and #3) show consistent separation of $\varepsilon_2(\omega)$ for different concentrations of contaminants. At 4.0 eV, $\varepsilon_2(\omega)$ for RDL 1 ranges between 2.34 – 2.53, whilst $\varepsilon_2(\omega)$ of RDL 3 is between 1.59 – 1.81. Furthermore, separation of $\varepsilon_2(\omega)$ for each contamination level is apparent at DUV for all batches. In addition, we perform correlation studies between the leakage current and optical properties of Cu RDL (Supplementary materials, Figure S5). Both $\varepsilon_1(\omega)$ and $\varepsilon_2(\omega)$ at 4.0 eV show excellent correlation with the measured $I_{leak}$. The correlation studies between $I_{leak}$ and spectral weight $W$ show an equally positive correlation, with $R^2$ reaching 0.89.

The experimental SE results and correlation studies of the test vehicles with complex microstructures show strong, positive correlation with metallic contaminants levels. As the metallic contaminants vary, however, they may penetrate less than 100 nm into the surface. Changes in the reflectance spectra in FDTD calculations also indicate that the optical properties of the contamination layer are dependent on the depth distribution of the contaminants. We model this behavior by simultaneously fitting both the thickness and complex dielectric function, using an iterative procedure previously conducted on spectroscopic ellipsometry studies on interfaces[21]. The fitting process relies on self-consistent parameter control to obtain a reproducible fit. Our fitting for RDL 1, RDL 2, and RDL 3 estimate the contamination layer thickness to be between 40 – 100 nm, in agreement with thickness estimation from XPS.

In conclusion, using a high-resolution spectroscopic ellipsometry (SE) supported with FDTD calculations, we demonstrate a remarkably fast, non-destructive method to probe various metallic contaminants at the surface and interface of complex polyimide heterostructures. The concentration of contaminants impacts the optical and electronic properties on the surface and are directly correlated to contact resistance ($R_C$) and leakage current ($I_{leak}$) in under-bump vias and redistribution layer. Spectroscopic ellipsometry is shown to be sensitive to the contamination level, where the complex dielectric function successfully identifies variations in minute contamination levels (1.3% – 30%) in thin films and complex interconnect structures. Furthermore, FDTD calculations provide qualitative insight into the reflectance spectra obtained from SE, revealing the spatial distribution and relative depth of metallic contaminants. Our result shows that the concentration, spatial distribution, and type of metallic contaminants determine the complex dielectric function of complex polyimide heterostructures. While SE can be very sensitive to metallic contaminants, improved theoretical approaches may be necessary to determine quantitatively metal contamination information from SE measured dielectric functions in the DUV.

## SUPPLEMENTARY MATERIAL

See supplementary material for discussions and plots concerning the experimental setup, the optical model used in ellipsometry, and *ex-situ* characterizations of the test vehicles used in this study.

## ACKNOWLEDGMENTS

This work is supported by AMAT-NUS Project (R-144-000-462-592), the Ministry of Education of Singapore (MOE) AcRF Tier-2 (MOE-T2EP50220-0018 and MOE-T2EP50122-0014), NRF - NUS Resilience and Growth Postdoctoral Fellowships (R-144-000-455-281 and R-144-000-459-281), and NUS Core Support (Grant No. C-380-003-003-001). The authors also thank the Singapore Synchrotron Light Source (SSLS) for providing the facility necessary for conducting the research. SSLS is a National Research Infrastructure under the Singapore National Research Foundation.

## AUTHOR DECLARATIONS

**Conflict of Interest**

The authors have no conflicts of interest to disclose.

# REFERENCES

[1] Harry Hedler, Thorsten Meyer, and Barbara Vasquez, (2004).

[2] Prayudi Lianto, Chin Wei Tan, Qi Jie Peng, Abdul Hakim Jumat, Xundong Dai, Khai Mum Peter Fung, Guan Huei See, Ser Choong Chong, Soon Wee David Ho, Siew Boon Serine Soh, Seow Huang Sharon Lim, Hung Ming Calvin Chua, Ahmad Abdillah Haron, Huan Ching Kenneth Lee, Mingsheng Zhang, Zhi Hao Ko, Ye Ko San, and Henry Leong, in *2020 IEEE 70th Electronic Components and Technology Conference (ECTC)* (2020), pp. 1126.

[3] Douglas C. H. Yu, in *2017 IEEE International Electron Devices Meeting (IEDM)* (2017), pp. 3.6.1.

[4] J. H. Lau, M. Li, D. Tian, N. Fan, E. Kuah, W. Kai, M. Li, J. Hao, Y. M. Cheung, Z. Li, K. H. Tan, R. Beica, T. Taylor, C. T. Ko, H. Yang, Y. H. Chen, S. P. Lim, N. C. Lee, J. Ran, C. Xi, K. S. Wee, and Q. Yong, IEEE Transactions on Components, Packaging and Manufacturing Technology **7** (10), 1729 (2017).

[5] Xin-Jiang Long, Jin-Tang Shang, Huang Tao, and Zhang Li, in *2017 IMAPS Nordic Conference on Microelectronics Packaging (NordPac)* (2017), pp. 109.

[6] Marvin Bernt, Paul Van Valkenburg, David Surdock, and Prayudi Lianto, International Symposium on Microelectronics **2018** (1), 000207 (2018).

[7] Prayudi Lianto, King-Jien Chui, Bharat Bhushan, H. M. Calvin Chua, Leijun Tang, B. S. S. Chandra Rao, Xin Wang, Ai Long Wu, Yu Gu, Guan Huei See, and Arvind Sundarrajan, IEEE Transactions on Components, Packaging and Manufacturing Technology **7** (10), 1592 (2017).

8       Guangyong Xu, D. E. Eastman, B. Lai, Z. Cai, I. McNulty, S. Frigo, I. C. Noyan, and C. K. Hu,  **94** (9), 6040 (2003).

9       N. G. Orji, M. Badaroglu, B. M. Barnes, C. Beitia, B. D. Bunday, U. Celano, R. J. Kline, M. Neisser, Y. Obeng, and A. E. Vladar,  Nat Electron **1**, 10.1038/s41928 (2018).

10      R. Joseph Kline, Daniel F. Sunday, Donald Windover, and Benjamin D. Bunday,  Journal of Micro/Nanolithography, MEMS, and MOEMS **16** (1) (2017).

11      K. Riedling, *Ellipsometry for Industrial Applications*. (Springer Vienna, 2012).

12      Vimal K. Kamineni, Pratibha Singh, LayWai Kong, John Hudnall, Jamal Qureshi, Chris Taylor, Andy Rudack, Sitaram Arkalgud, and Alain C. Diebold,  Thin Solid Films **519** (9), 2924 (2011).

13      Alain C. Diebold, Florence J. Nelson, and Vimal K. Kamineni, in *Ellipsometry at the Nanoscale*, edited by Maria Losurdo and Kurt Hingerl (Springer Berlin Heidelberg, Berlin, Heidelberg, 2013), pp. 557.

14      Sonal Dey, Alain Diebold, Nick Keller, and Madhulika Korde, *Mueller matrix spectroscopic ellipsometry based scatterometry simulations of Si and Si/SixGe1-x/Si/SixGe1-x/Si fins for sub-7nm node gate-all-around transistor metrology*. (SPIE, 2018).

15      Parker Huang, YiYen Liu, Jay Chao, Chun Hung Lu, Stephen Chen, Jay Chen, Fei Shen, Jian Ding, Priya Mukundhan, and Timothy Kryman,  Microelectronic Engineering **137**, 111 (2015).

16      T. J. Whitcher, Angga Dito Fauzi, D. Caozheng, X. Chi, A. Syahroni, T. C. Asmara, M. B. H. Breese, A. H. Castro Neto, A. T. S. Wee, M. Aziz Majidi, and A. Rusydi,  Nat. Commun. **12** (1), 6980 (2021).

17   X. J. Yu, C. Z. Diao, T. Venkatesan, M. B. H. Breese, and A. Rusydi,  Review of Scientific Instruments **89** (11), 113113 (2018).

18   G. E. Jellison Jr. and F. A. Modine,  Applied Physics Letters **69** (3), 371 (1996).

19   B. Johs, C. M. Herzinger, J. H. Dinan, A. Cornfeld, and J. D. Benson,  Thin Solid Films **313-314**, 137 (1998).

20   H. Fujiwara, *Spectroscopic Ellipsometry: Principles and Applications*. (Wiley, 2007).

21   T. C. Asmara, I. Santoso, and A. Rusydi,  Review of Scientific Instruments **85** (12), 123116 (2014).

22   E.D. Palik, *Handbook of Optical Constants of Solids*. (Elsevier Science, 1998).

23   P. B. Johnson and R. W. Christy,  Phys. Rev. B, **9** (12), 5056 (1974).

24   Shaista Babar and J. H. Weaver,  Appl. Opt. **54** (3), 477 (2015).